\documentclass[11pt,a4paper]{article}

\usepackage{graphicx}
\usepackage{dcolumn}
\usepackage{bm}
\expandafter\let\csname equation*\endcsname\relax
\expandafter\let\csname endequation*\endcsname\relax
\usepackage{amsmath}
\usepackage{authblk}

\begin{document}

\title{X-type vortex and its effect on beam shaping}

\author[a]{Xiaoyan Pang \thanks{corresponding author:xypang@nwpu.edu.cn}} 
\author[a]{Weiwei Xiao}
\author[a]{Han Zhang}
\author[a]{Chen Feng}
\author[b]{Xinying Zhao\thanks{corresponding author:zhaoxinying@snnu.edu.cn}}

\affil[a]{$^1$ School of Electronics and Information, Northwestern Polytechnical University, Xi'an, China}
\affil[b]{$^2$ School of Physics and Information Technology, Shaanxi Normal University, Xi'an, China}



\begin{abstract}
In this article we propose a new type of optical vortex, the X-type vortex.
This vortex inherits and develops the conventional noncanonical vortex, 
i.e., it no longer has a constant phase gradient around the center, while the intensity keeps invariant azimuthally. 
The strongly focusing properties of the X-type vortex and its effect on the beam shaping in three-dimensional (3D) fields are analyzed.
 The interesting phenomena, which cannot be seen in canonical vortices, are observed, for instance the ‘switch effect’ which shows that the intensity pattern can switch from one transverse axis to another in the focal plane by controlling the phase gradient parameter. 
It is shown that by adjusting the phase gradient of this vortex, the focal field can have marvelous patterns, from the doughnut shape to the shapes with different lobes, and the beam along propagation direction will form a twisting shape in 3D space with controllable rotation direction and location.
The physical mechanisms underlying the rule of the beam shaping are also discussed, which generally say that the phase gradient of the X-type vortex, the orbital angular momentum, the polarization and the `nongeneric' characteristic contribute differently in shaping fields.
This new type of vortex may supply a new freedom for tailoring 3D optical fields,
 and our work will pave a way for exploration of new vortices and their applications.
\end{abstract}

\maketitle
\section{Introduction}
Vortices, also called phase singularities, have been studied more than 30 years in optics \cite{s2001,gbur2017,vor19}.
Since their topological structure and carrying orbital angular momentum (OAM),
optical vortices are one of the main objects of study in singular optics, 
and also have found numerous applications, ranging from optical tweezers \cite{AP2}, microscopy \cite{Ph2013} and imaging \cite{im2006}, optical communications \cite{W15}, to metrology \cite{Xie17} and astronomy\cite{Sw08}.
Although the vortex is a natural structure and can occur generically in light fields \cite{s2001,gbur2017},
the type of the vortices in 
most studies and applications is only the `canonical' one \cite{vor19}, i.e. the phase gradient is constant along a circular path around the vortex center, 
which can be simply expressed as $A e^{{\rm i}n\phi}$ ($A$ is the field amplitude, $\phi$ is azimuthal angle and $n$ denotes the topological charge).
Essentially, the $2\pi n$ of phase change does not need to be distributed uniformly around the vortex center;
in other words, the vortex can be `noncanonical' \cite{Fr1993}. 
The `noncanonical' means that the phase gradient of the vortex is not constant, but the topological charge is the same as it in its `canonical' counterpart \cite{Sch96,Mol01,Roux04}.  

The noncanonical vortices actually were studied by several scientists around the year 2000 \cite{Sch96,Mol01,Roux04,Fr97,Kim03},
while only the forms of the noncanonical vortices satisfying the paraxial wave equation were taken into consideration.
There, the `noncanonical' vortices were also called `anisotropic' vortices \cite{Fr97,Kim03}, since not only the phase but also the amplitude of such vortices were spatially anisotropic. 
Perhaps this anisotropic structure limited the study of these vortices theoretically and experimentally, 
so the noncanonical vortices were visited rarely in recent years \cite{Maji17}.
In this article, we will propose a new type of noncanonical vortices which is easy to construct and its focusing properties  will also be discussed.

The application/processing of laser is always accompanied by beam shaping, from  reshaping the `elliptical' far-field pattern into almost `circular' at an early stage \cite{Braat1995},  to generating  a flat-top focus \cite{Nodop19} and multi-foci profiles \cite{Guo11} in a two-dimensional (2D) plane, and shaping special structures in three-dimensional (3D) fields \cite{2006Three,Chang17,Por2020}, such as  twisting beams \cite{Daria11}, optical cages and needles \cite{Weng17,man2018,Chen21}, etc.
The beam shaping is still an active research field that supplies various methods in  inverse problems for getting desired beam patterns \cite{Chang17,Por2020,2014Light} and also gives rise to new research directions in study of structured light \cite{Otte2018,forbes2021}. 
There are many ways used in beam shaping, basically, including tailoring single parameter, like controlling the spatial distribution of the phase \cite{Nodop19,Zhao18},  the polarization \cite{2006Three,man2018}, the coherence \cite{Zhao19,Tong20} of the input beams, and manipulating multi-parameters, for instance the phase-coherence method \cite{Hua2019,2020Shaping}, the phase-polarization method \cite{2014Light}, the amplitude-phase-polarization joint method \cite{hao2014}.  
Here we will show that by controlling the strength factor of the proposed noncanonical vortex, the focal pattern and the beam shape in a 3D focused fields can experience marvelous changes.
That will provide a new perspective for beam shaping and may be used in many applications, such as optical tweezers \cite{vor19,AP2} and laser material processing \cite{2019Tailored}.

A conventional noncanonical vortex embedded in a beam center can be expressed as \cite{Sch96,Mol01}
\begin{equation}\label{eq1}
E(x,y)=A(x,y)(x+{\rm i}\sigma_c y)^n,
\end{equation}
with $A(x,y)$ the transverse distribution of the host beam, and the topological charge of the vortex is  $\pm n$ ($n>0$), i.e. $+n$ for ${\rm Re}[\sigma_c]>0$, and $-n$ for ${\rm Re}[\sigma_c]<0$.
$\sigma_c$  determining the noncanonical strength of the vortex is usually called `anisotropy parameter'.
This equation also indicates that the topological charge of this noncanonical vortex is independent of $|\sigma_c|$ if ${\rm Re}[\sigma_c]\neq0$. 
When Eq. (\ref{eq1}) is written in the polar coordinates $(\rho,\phi)$ ($x=\rho\cos\phi$, $y=\rho\sin\phi$) as 
\begin{align}\label{eq2}
E(\rho,\phi)&=A(\rho,\phi)(\rho\cos\phi+{\rm i}\sigma_c\rho\sin\phi)^n \nonumber \\
&=A(\rho,\phi)\rho^n M(\phi,\sigma_c) e^{{\rm i}n\arctan(\sigma_c\tan \phi)},
\end{align}
with $M(\phi,\sigma_c)=\left(\sqrt{\cos^2\phi+\sigma^2_c\sin^2 \phi} \right)^n$,
one can see clearly that as $|\sigma_c|$ varies, except the topological charge, both the phase and the amplitude change their distributions,
which means that the field with this vortex is `anisotropic'  on both the phase and the amplitude.
In other words, the `anisotropic' distributions of the phase and amplitude are coupling in the `anisotropic parameter' $\sigma_c$. 
 That makes the noncanonical vortex not a `pure' description of the helical wavefront. 
Thus, here we propose a new type of `pure' noncanonical vortex (whose amplitude is independent of the `anisotropy parameter' $\sigma_c$) named as {\em X-type vortex} (the reason for this name will be explained later).
This X-type vortex can be expressed as
\begin{equation}\label{eq3}
E^{(X)}(\rho,\phi)=A(\rho,\phi)\rho^n e^{{\rm i}n\arctan(\sigma_c\tan \phi)},
\end{equation}
which shows that $\sigma_c$ only determines the phase structure
(here we still adopt the convention that $\sigma_c$ is named as `anisotropic' parameter).
When $\sigma_c=\pm 1$, this vortex becomes a $\rho$ type canonical vortex and
if $A(\rho,\phi)$ is also Gaussian Eq.~(\ref{eq3}) is actually an approximation expression of a LG beam in the transverse plane.
For simplicity but without loss of generality,
from here on, we assume $A(\rho,\phi)=e^{-\rho^2/w^2_0}$ (i.e. a Gaussian profile), with $w_0$ the beam width.
This proposed X-type vortex can be generated easily through the common methods for the canonical vortex, such as by using a spatial light modulator or a  spatial phase plate.

From Eq.~(\ref{eq3}), we can see that the amplitude or the intensity gradient of this vortex along the $\phi$ direction is zero, while the phase $\Phi=n\arctan(\sigma_c\tan \phi)$ is a nonlinear function of $\phi$ with its gradient calculated as
\begin{equation}\label{eq4}
\nabla\Phi=\frac{{\rm d}\Phi}{{\rm d}\phi}=\frac{n \sigma_c}{\cos^2\phi+\sigma^2_c\sin^2\phi}.
\end{equation}
The phase and the intensity distribution of the X-type vortex are shown in Fig.~\ref{Fig1}, where the topological charge is $1$ in the first two lines and is $2$ in the last two lines. $\sigma_c$ are selected as $0.5$, $1$ and $2$ separately in the three columns.  
Then based on Eqs. (\ref{eq3}) and (\ref{eq4}) and Fig. \ref{Fig1}, the properties of the X-type vortex can be summarized as follows:
\begin{figure}[ht]
	\centering
	\includegraphics[width=8.0cm]{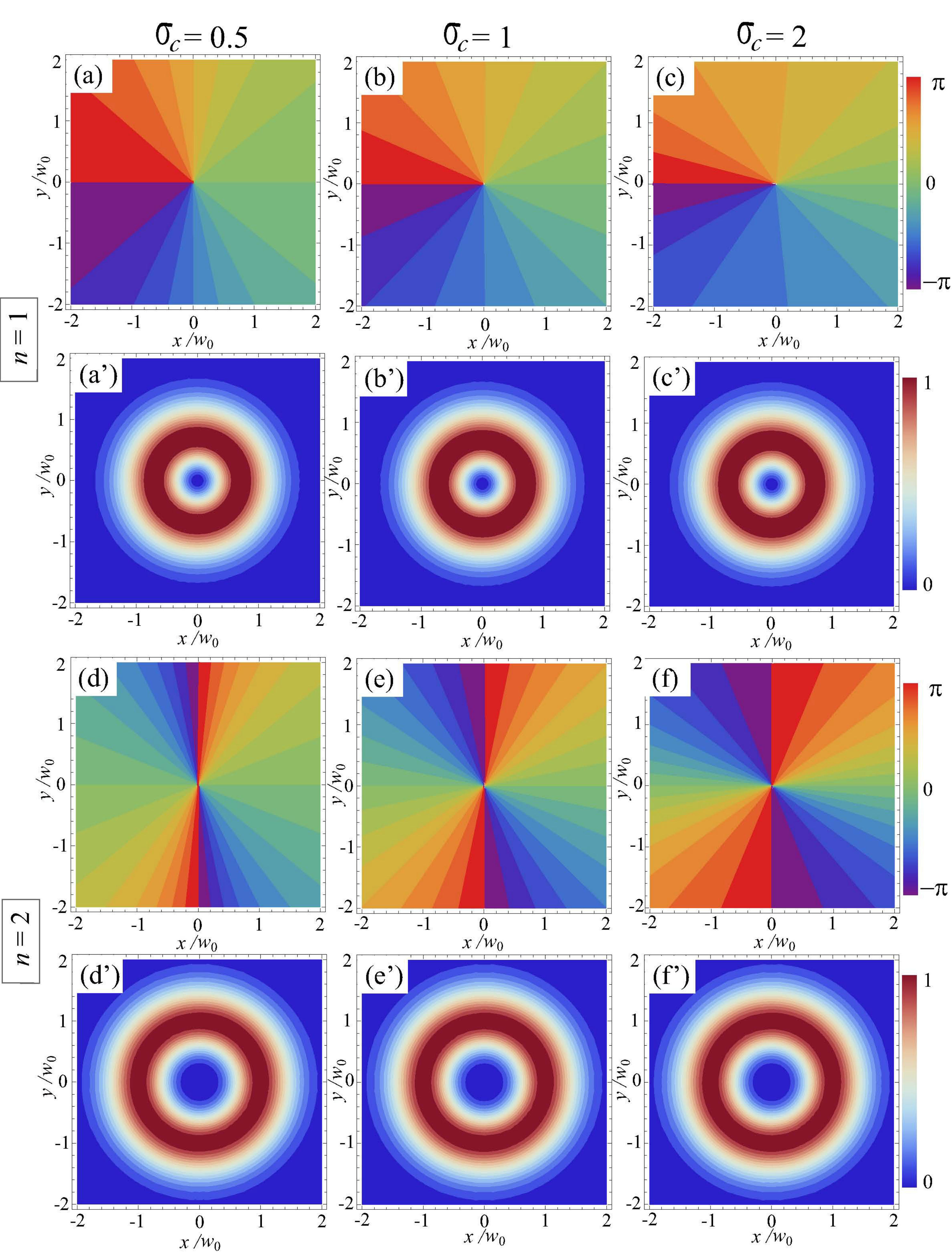}
	\caption{Color-coded plots of the phases (a)-(f) and the intensity (a')-(f') of X-type vortex. In the plots of the first two lines $n=1$, while in the last two lines $n=2$. $\sigma_c=0.5, 1, 2$ for the first, second and third columns.}
	\label{Fig1}
\end{figure}
\begin{itemize}
\item The intensity distribution is independent of $\sigma_c$, see Fig.~\ref{Fig1} (a')-(f').
\item $\nabla\Phi(\phi=0)=\nabla\Phi(\phi=\pi)$ and $\nabla\Phi(\phi=\pi/2)=\nabla\Phi(\phi=-\pi/2)$. This means that the phase gradient of this vortex is symmetric about both the $x$-axis and the $y$-axis, i.e. it has a `cross' (`X') symmetry, so this is the reason for calling this vortex as the X-type vortex [see Fig.~\ref{Fig1} (a)-(f)].
\item If $|\sigma_c|=1$, $\nabla\Phi=\pm n$, i.e. the phase gradient is a constant.
In this case, the X-type (noncanonical) vortex degenerates into a canonical vortex [see Fig.~\ref{Fig1} (b) and (e)].
\item If $|\sigma_c|<1$, the phase gradient gets its maximum $|\nabla\Phi|=|n/\sigma_c|$ at $\phi=\pm\pi/2$, the $y$-axis.
Also the phase changes slow near the $x$-axis and fast near the $y$-axis, i.e. the slope of the phase is steeper around the $y$-axis [see Fig.~\ref{Fig1} (a) and (d)].
	\item If $|\sigma_c|>1$, the phase gradient arrives at its peak $|\nabla\Phi|=|n\sigma_c|$ at $\phi=0,\pi$, the $x$-axis.
In this situation, the phase changes fast near the $x$-axis and slow near the $y$-axis, i.e. the slope of the phase is steeper around the $x$-axis  [see Fig.~\ref{Fig1} (c) and (f)].
\end{itemize}


\section{Focused field}
\begin{figure}[ht]
	\centering
	\includegraphics[width=8.0cm]{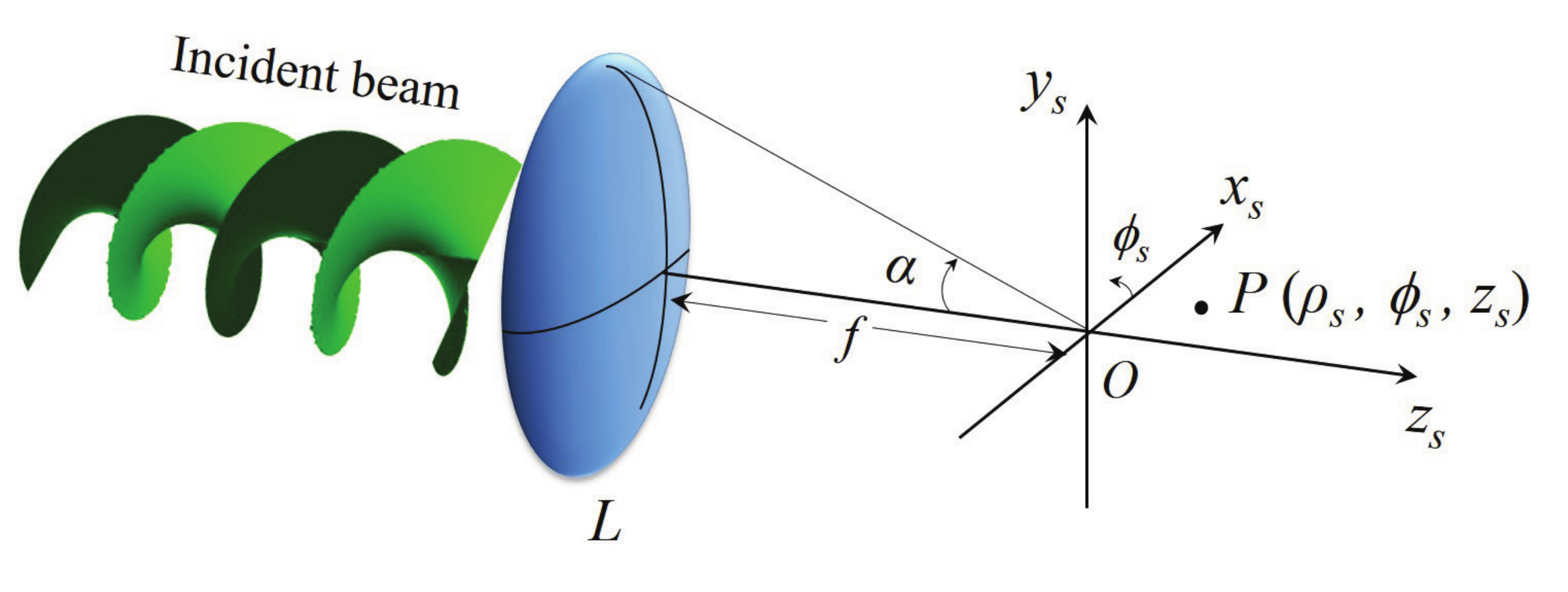}
	\caption{A strongly focusing system with a vortex beam as the incident field.}
	\label{Fig2}
\end{figure}

In this section we consider an aplanatic, high numerical aperture (NA) focusing system with semi-aperture angle $\alpha$ and focal length $f$,   where  the beam with the X-type vortex is the incident field, see Fig.~\ref{Fig2}. 
Assume that the incident field is linearly polarized at $x$ direction and its complex field at the entrance plane is expressed as Eq. (\ref{eq3}), i.e. a Gaussian type noncanonical vortex with $x$ polarization,
then according to the Richards-Wolf vectorial diffraction theory \cite{RichardWolf} the electric field ${\bf E}^f$ in the focal region at an observation point $P(\rho_{s},\phi_{s},z_{s})$ can be expressed  as
\begin{equation}\label{eq5}
\begin{split}
 &{\bf E}^f(\rho_{s},\phi_{s},z_{s}) =
\left[ 
\begin{matrix}
e_{x}\\ e_{y}\\ e_{z}
\end{matrix}  
\right] \\
 &=-\frac{{\rm i}kf}{2\pi}\int_{0}^{\alpha}\int_{0}^{2\pi} E^{X}(r,\phi)\sqrt{\cos\theta}\sin\theta e^{{\rm i}kz_{s}\cos\theta}  \\
&\times \left[
\begin{matrix}   
\cos\theta+\sin^2 \phi(1-\cos\theta)  \\
(\cos\theta-1)\cos \phi \sin \phi \\
-\sin\theta \cos \phi  \\
\end{matrix}      
\right] e^{{\rm i}k\rho_{s}\sin\theta\cos(\phi-\phi_{s})} 
{\rm d}\phi{\rm d}\theta,               
\end{split}
\end{equation}
where $k$ is the wave number $k=2\pi/\lambda$ ($\lambda$ is the wavelength) and $(\rho_{s},\phi_{s},z_{s})$ are the cylindrical coordinates in image space ($\rho_{s}= \sqrt{x_{x}^2+y_{s}^2}$).
Here $E^{X}(r,\phi)=E^{X}(\rho,\phi)$ by applying the Abbe sine condition $\rho=r\sin\theta$ in the aplanatic system.

From Eq.~(\ref{eq5}), we note the following relations existing for the fields components $e_x$, $e_y$ and $e_z$:
\begin{align}
\label{es1}&e^*_x(\rho_{s},\phi_{s},z_{s})=-e_x(\rho_{s},\pi-\phi_{s},-z_{s}),  \\
\label{es2}&e^*_y(\rho_{s},\phi_{s},z_{s})=+e_y(\rho_{s},\pi-\phi_{s},-z_{s}), \\
\label{es3}&e^*_z(\rho_{s},\phi_{s},z_{s})=-e_z(\rho_{s},\pi-\phi_{s},-z_{s}). 
\end{align}

When $\sigma_c=1$, i.e. the noncanonical case, besides of the relations above, the field components also satisfy:
\begin{align}
\label{es4}&e_x(\rho_{s},\phi_{s},z_{s})=(-1)^n e_x(\rho_{s},\pi+\phi_{s},z_{s}),  \\
\label{es5}&e_y(\rho_{s},\phi_{s},z_{s})=(-1)^n e_y(\rho_{s},\pi+\phi_{s},z_{s}), \\
\label{es6}&e_z(\rho_{s},\phi_{s},z_{s})=-(-1)^n e_z(\rho_{s},\pi+\phi_{s},z_{s}). 
\end{align}

\subsection{$n=1$}
In this section, we will examine the focusing properties of the first order X-type vortex  ($n=1$), and discuss its role in beam shaping.

\begin{figure}[ht]
	\centering
	\includegraphics[width=8.0cm]{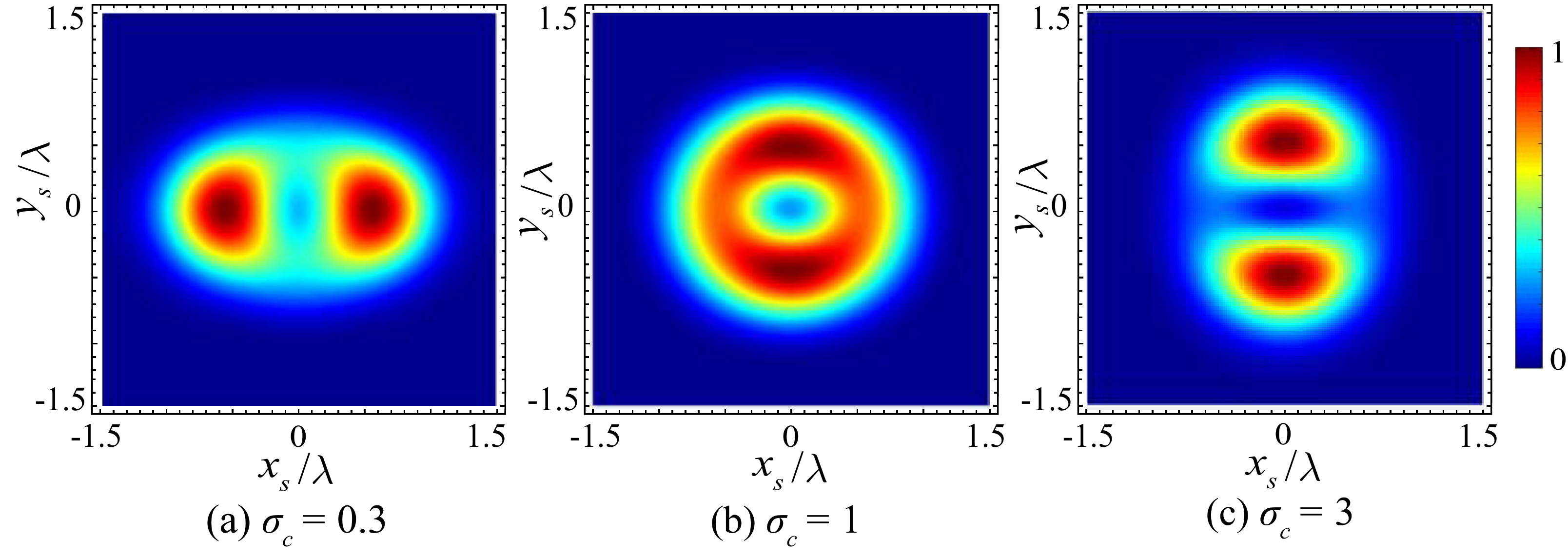}
	\caption{Intensity distribution on the focal plane: (a) $\sigma_c=0.3$, (b) $\sigma_c=1$, (c) $\sigma_c=3$. In all the panels $n=1$, $\alpha=60^\circ$.}
	\label{Fig3}
\end{figure}

\begin{figure}[ht]
	\centering
	\includegraphics[width=7.5cm]{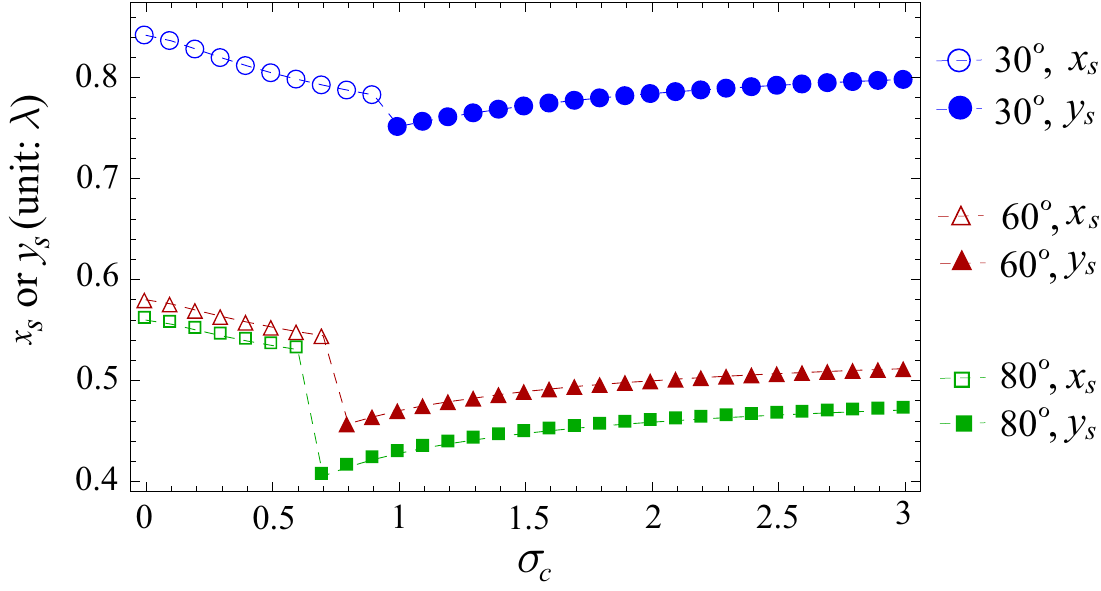}
	\caption{Locations of the intensity maxima on the focal plane with the anisotropic parameter $\sigma_c$. Here the semi-aperture angle is selected as  $30^\circ$ (in blue), $60^\circ$ (in red) and $80^\circ$ (in green).}
	\label{Fig4}
\end{figure}

Firstly, we will look at the field at the focal plane ($z_s=0$).
The intensity distribution of the total field is shown in Fig.~\ref{Fig3}, where $\sigma_c$ is selected as $0.3$ [plot (a)], $1$[plot (b)] and $3$ [plot (c)], and the semi-aperture angle $\alpha=60^\circ$.
As we can see, the anisotropy parameter $\sigma_c$ has a strong effect on the intensity of the field,
i.e., the intensity  pattern  has two lobes on the $x_s$-axis  for $\sigma_c=0.3$, two lobes on the $y_s$-axis for  $\sigma_c=3$ and a (uneven) doughnut shape for $\sigma_c=1$.
In order to see this phenomenon more directly, the position variation of the intensity maxima with $\sigma_c$ is drawn in Fig.~\ref{Fig4}, where the hollow marker denotes the position on $x_s$-axis and the solid marker denotes that on the $y_s$-axis.
Note that since for each case there are two maxima in the focal plane and they are symmetrical about the origin, here only one of them is shown.
 
By analyzing Figs.~\ref{Fig3} and \ref{Fig4}, we can get:
\begin{itemize}
\item Generally, when $\sigma_c>1$, the intensity maxima are located on the $y_s$-axis, while when $\sigma_c<1$ (a little bit far from $1$) the intensity maxima are on the $x_s$-axis, and when $\sigma_c$ is near $1$ (slightly $<1$), the intensity  will have a doughnut shape. 
\item There exists a switch value $\sigma^{sw}_c$, and when $\sigma_c$ passes $\sigma^{sw}_c$ the intensity maxima will move from $x_s$-axis to $y_s$-axis (or vice versa). 
\item As the semi-aperture $\alpha$ increases, on one hand the intensity maxima will be closer to the central axis (i.e. the distance between two intensity maxima is shorter), and on the other hand the  switch value $\sigma^{sw}_c$ will become smaller. 
\end{itemize}

This focal shaping phenomenon physically can be explained by the joint effect of the phase gradient $\nabla\Phi$, the polarization of the incident beam and the strongly focusing. 
As we discussed in section 2, when $|\sigma_c|<1$, $\nabla\Phi=1/|\sigma_c|$ (the maximum) on the $y$-axis, while $\nabla\Phi=|\sigma_c|$ on the $x$-axis for  $|\sigma_c|>1$. 
This non-unit value of $\sigma_c$ causes the `polar' intensity distribution on these two axes (stronger on $y_s$ for $|\sigma_c|<1$, on $x_s$ for $|\sigma_c|>1$) in the far field (or in the focused field)   as the beam is diffracted, and this `polar'  distribution will be strengthened by the high NA system. 
Therefore the two-lobes patterns of the intensity for $\sigma_c<1$ and $\sigma_c>1$  are formed [like Figs.~\ref{Fig3} (a) and (c)].
The polarization of the incident beam is linear, $x$-polarization, which will lead to the intensity distributed more along the $y_s$-axis when the beam is strongly focused \cite{RichardWolf}, and as the semi-aperture angle is larger, this anisotropic distribution becomes stronger. 
That is the reason for the uneven doughnut shape in the case of $\sigma_c=1$ [Fig.~\ref{Fig3}(b)] and the smaller value of $\sigma^{sw}_c$ for the larger $\alpha$ (see the curves for $60^\circ$ and $80^\circ$ in Fig.~\ref{Fig4}). 
 
Since the same incident intensity distribution can lead to distinguishing intensity patterns in the focal plane, there may exist intensity evolutions along the beam propagation.

\begin{figure}[ht]
	\centering
	\includegraphics[width=8.0cm]{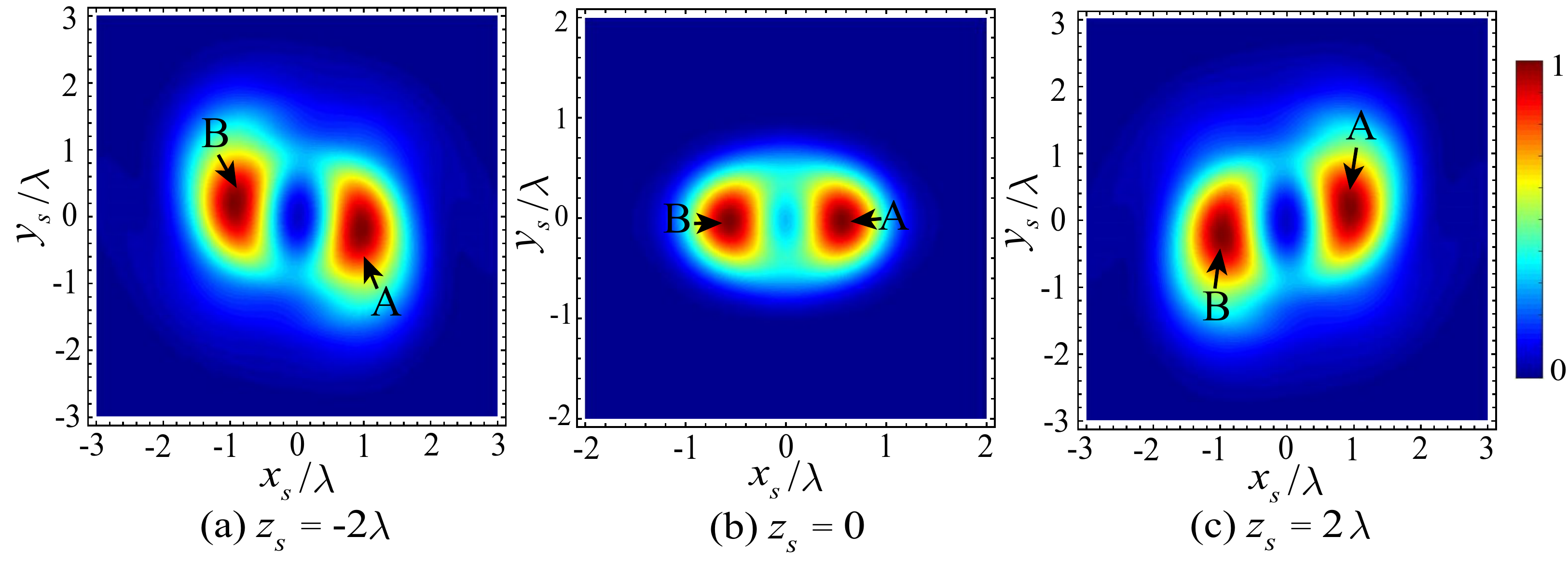}
	\caption{Intensity distribution on the transverse planes along the propagation axis. Here
$\alpha=60^\circ$, $\sigma_c=0.3$, $n=1$.}
	\label{Fig5}
\end{figure}

Secondly, we will discuss the intensity variation with the beam  propagation.
Fig.~\ref{Fig5} shows the intensity distribution on the transverse planes at $z_s=-2\lambda$, $0$ and $+2\lambda$, where $\alpha=60^\circ$ and $\sigma_c=0.3$.
We can see two intensity maxima (denoted by `A' and `B') still existing in different transverse planes, 
but they rotate with the beam propagation.

\begin{figure}[ht]
	\centering
	\includegraphics[width=7.5cm]{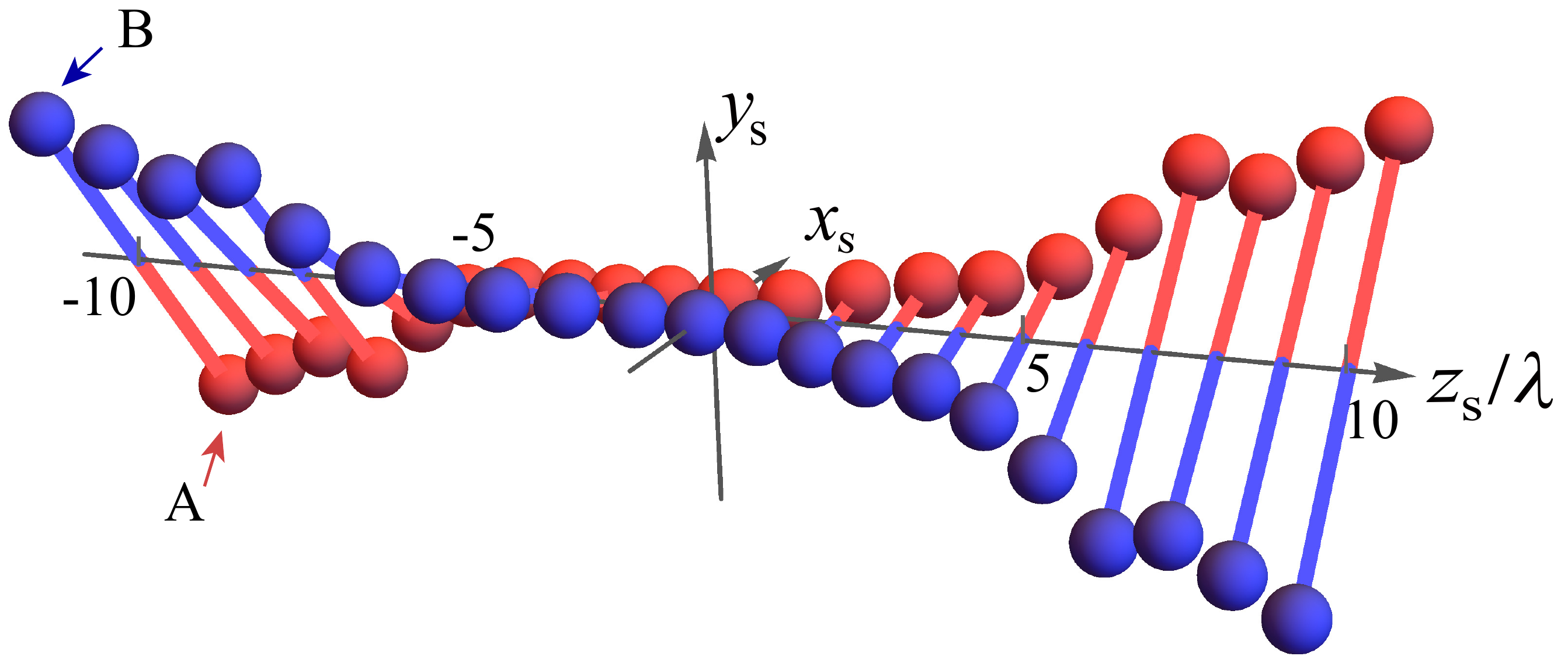}
	\caption{Positions of the two intensity maxima with $z_s$ from $-10\lambda$ to $10\lambda$ (see \textbf{Video 1}). Here
$\alpha=60^\circ$, $\sigma_c=0.3$, $n=1$. }
	\label{Fig6}
\end{figure}

\begin{figure}[h]
	\centering
	\includegraphics[width=8.0cm]{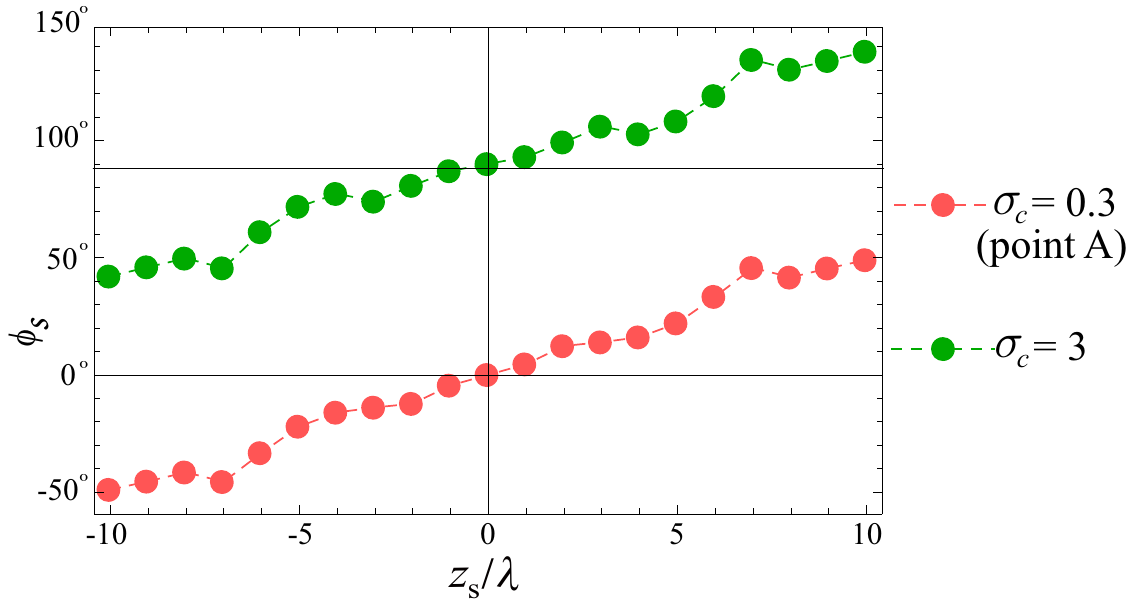}
	\caption{Azimuthal angle of the intensity maxima along the propagation direction for the incident beam of charge $+1$. Here $\alpha=60^\circ$. }
	\label{Fig7}
\end{figure}

This rotation can be seen more straightly in Fig.~\ref{Fig6}, where the positions of the intensity maxima (points A and B, represented by red and blue spheres respectively) are drawn with $z_s$ from $-10\lambda$ to $+10\lambda$.
It can be seen that the intensity maxima rotate in a counter-clockwise manner along the propagation direction.
The azimuthal angle of one maximum (point A, denoted by red circles) is plotted in Fig.~\ref{Fig7}, from which we can get in the distance $-10\lambda\leq z_s\leq +10\lambda$ the maximum point rotating from about $-50^\circ$ to $+50^\circ$. 
The case for $\sigma_c=3$ is also drawn in Fig.~\ref{Fig7}, and it shows that although there are about $90^\circ$ difference between the  maxima positions for $\sigma_c=0.3$ and $\sigma_c=3$, the maximum point for $\sigma_c=3$ also rotates in the same manner and has the same total rotation angle as it does for $\sigma_c=0.3$.
It is also easy to illustrate that this beam rotation only occurs when $|\sigma_c|\neq 1$.
Actually, the symmetry relations have implied the possibility of this rotation.
From Eqs.~(\ref{es1})-(\ref{es3}), we can get
\begin{equation}\label{es}
|e_j(\rho_{s},\phi_{s},z_{s})|=|e_j(\rho_{s},\pi-\phi_{s},-z_{s})|,  ~~(j=x,y,z)
\end{equation}
which means that there may exist an azimuthal rotation of the intensity distributions between the field at two sides of the origin. 
As we discussed before, the non-unit value of $|\sigma_c|$ ($|\sigma_c|\neq 1$) causes the `polar'  intensity distribution of the field when the beam passes through a lens. Thus the rotation of the beam pattern can be seen.

\begin{figure}[h]
	\centering
	\includegraphics[width=8.0cm]{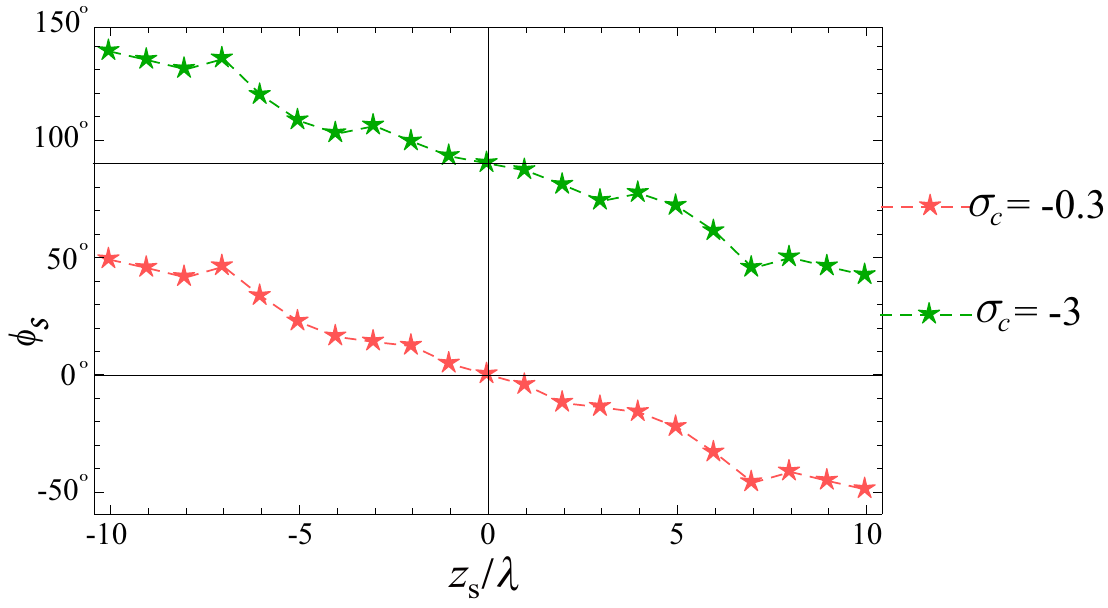}
	\caption{Azimuthal angle of the intensity maxima along the propagation direction for the incident beam of charge $-1$. Here $\alpha=60^\circ$. }
	\label{Fig8}
\end{figure}

This beam shaping, the rotation of the beam pattern, physically may be caused by the OAM of the noncanonical vortex.
When $\sigma_c<1$, for instance $\sigma_c=-0.3$, the topological charge of the incident beam becomes $-1$, which means that the OAM has the opposite orientation of that in $\sigma_c=0.3$.
As it is shown in Fig.~\ref{Fig8}, the azimuthal angle of the maximum point for $\sigma_c=-0.3$ varies from around $50^\circ$ to $-50^\circ$ along the propagation direction, i.e. the beam pattern rotates in a clockwise manner, which is opposite to that of $\sigma_c=0.3$.
This opposite rotation also can be seen  for $\sigma_c=-3$  in Fig.~\ref{Fig8}.
By comparing these two figures, Figs.~\ref{Fig7} and \ref{Fig8}, we can get the conclusion: 
for $\sigma_c>0$, the beam pattern will rotate counter-clockwise; whereas  for $\sigma_c<0$, it will rotate clockwise.
(Note that the rotation of the beam will not be observed if $|\sigma_c|=1$. )

\subsection{$n>1$}
In this section, we will consider  the X-type vortex with higher topological charge, i.e. $n>1$.
The vortex with $n=2$ and its role in the beam shaping is discussed in detail at the first two parts,
then the results are generalized to other X-type vortices with higher charge.

\begin{figure}[ht]
	\centering
	\includegraphics[width=8.0cm]{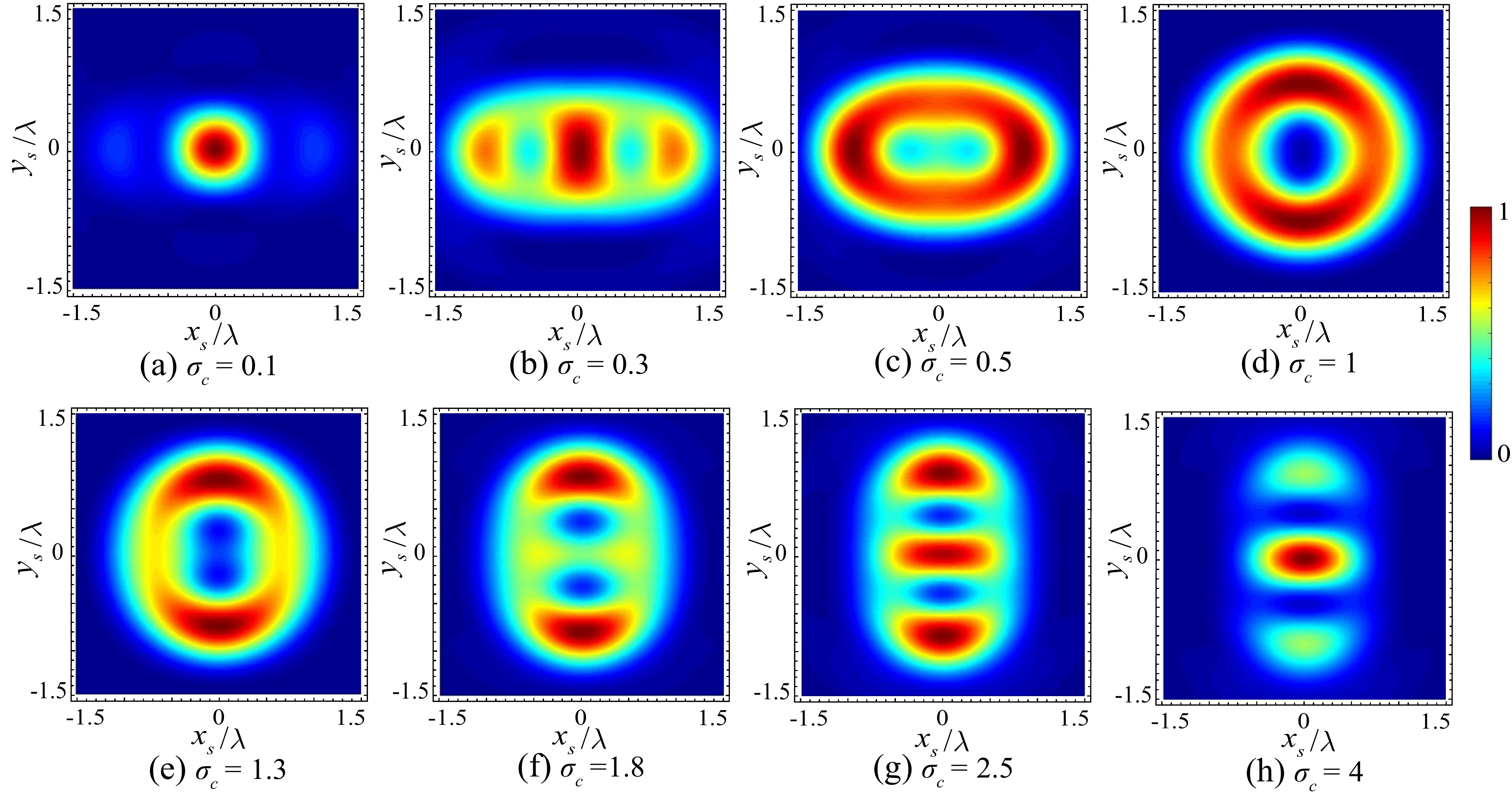}
	\caption{Intensity distribution on the focal plane: $\sigma_c=0.1$(a), $0.3$(b), $0.5$(c), $1$(d), $1.3$(e), $1.8$(f), $2.5$(g) and $4$(h). In all the panels $n=2$, $\alpha=60^\circ$. }
	\label{Fig9}
\end{figure}

\begin{figure}[ht]
	\centering
	\includegraphics[width=8.0cm]{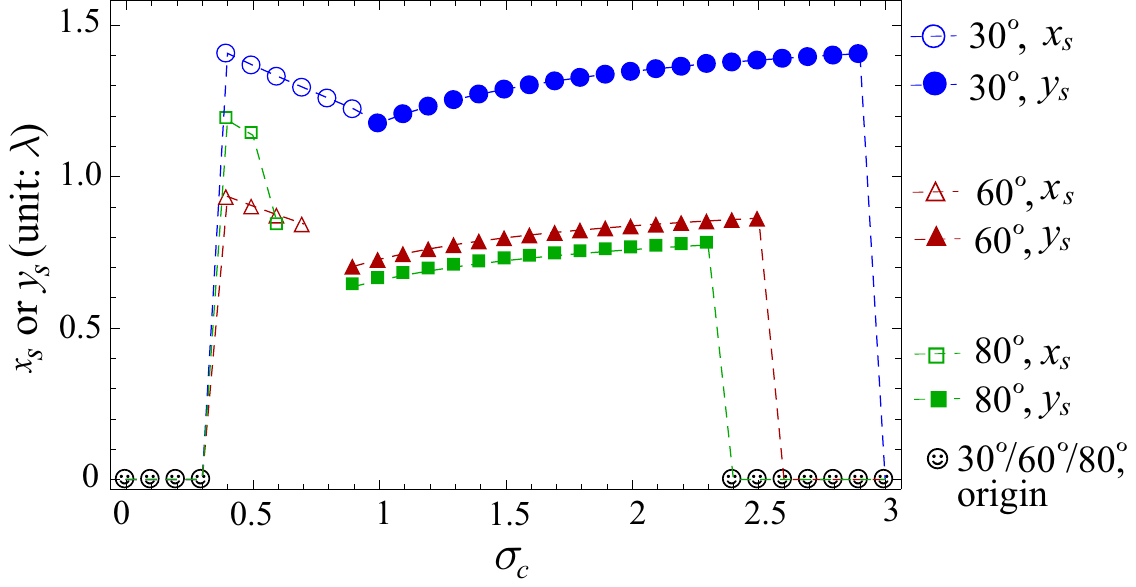}
	\caption{Locations of the intensity maxima on the focal plane with the anisotropic parameter $\sigma_c$. Here the semi-aperture angle is selected as  $30^\circ$ (in blue), $60^\circ$ (in red) and $80^\circ$ (in green), and $n=2$. }
	\label{Fig10}
\end{figure}

Firstly, let us analyze the field at the focal plane for $n=2$.
The intensity distribution for different values of $\sigma_c$ is shown in Fig.~\ref{Fig9}, where $\alpha=60^\circ$.
It is interesting to see that the intensity at the focal plane has `rich' patterns: 
 a two-lobe shape and a (uneven) doughnut shape (the similar patterns in $n=1$); 
one bright core at center, a core with two `wings' (the `wings' can be spread along the $x_s$-axis or the $y_s$-axis).
So there can exist one intensity maximum, two or three intensity maxima for $n=2$.
The positions of the intensity maxima with the variation of $\sigma_c$ are displayed in Fig.~\ref{Fig10}, where as it is done for $n=1$ (Fig.~\ref{Fig4}), only one maxima is drawn when there coexist two symmetric maxima. Note in this figure, the discontinues on the curves for $\alpha=60^\circ$ and $\alpha=80^\circ$ mean that when  $\sigma_c$ gets these values, the maxima are not located at the $x_s$- or $y_s$-axis.

By comparing Figs.~\ref{Fig9},\ref{Fig10} with Figs.~\ref{Fig3}, \ref{Fig4}, we can get:
\begin{itemize}
\item Generally, the intensity maxima have a tendency to stay on the $y_s$-axis for $\sigma_c>1$ and  on the $x_s$ axis for $\sigma_c<1$, which is similar to the trend of the maxima when $n=1$. While as $\sigma_c$ goes far from $1$ (like $\sigma_c=0$ or $3$), the intensity maxima will move to the origin (i.e. the position of the geometrical focus). 
\item The `switch effect' also exists, which means that the intensity maxima will move from $x_s$ axis to $y_s$ axis (or vice versa) once $\sigma_c$ arrives at $\sigma^{sw}_c$, however here $\sigma^{sw}_c$ can be a short interval rather than one number, for instance when $\alpha=60^\circ$, $\sigma_c^{sw}$ is about $0.7\sim 0.9$ (see the curve for $60^\circ$ in Fig.~\ref{Fig10}).
\item When the semi-aperture $\alpha$ increases, the same behavior in $n=1$ can also be found: the intensity maxima (except those maxima locating at the focus) are closer to the central axis and $\sigma^{sw}_c$ becomes smaller, while the difference in $n=2$ is that the maxima more likely move to the focus (i.e. the maximum goes back to the focus when $\sigma_c=2.4$  for $\alpha=80^\circ$, and $\sigma_c=2.9$ for $\alpha=30^\circ$.)
\end{itemize}

The difference appeared in the focal shaping for $n=2$ is brought by the `non-generic' characteristic of the higher order X-type vortex.
When $n>1$, the phase singularity (i.e. the vortex) in the incident vortex field is no longer fundamental, and this high order singularity will  decompose into $n$ singularities with charge $\pm 1$ during the beam propagation according to the conservation law of the topological charge \cite{F1994}.
At the locations of the phase singularities, the intensity of the corresponding field component is null, thus the total intensity pattern is re-distributed.

\begin{figure}[ht]
	\centering
	\includegraphics[width=8.0cm]{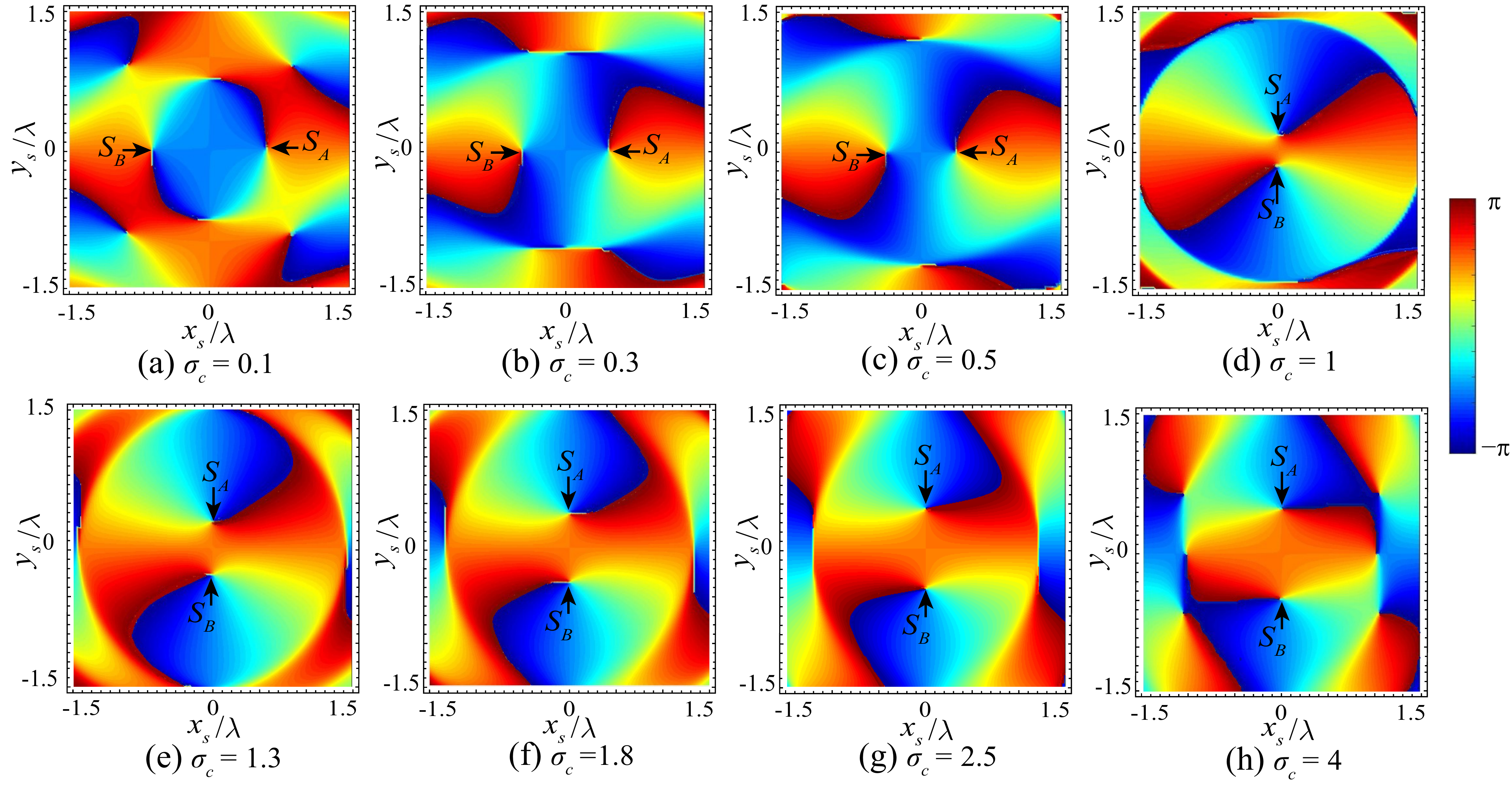}
	\caption{Phase distribution of $e_x$ on the focal plane:  $\sigma_c=0.1$(a), $0.3$(b), $0.5$(c), $1$(d), $1.3$(e), $1.8$(f), $2.5$(g) and $4$(h). In all the panels $n=2$, $\alpha=60^\circ$. }
	\label{Fig11}
\end{figure}

The  phase structure of the $e_x$ component  corresponding to the field in Fig.~\ref{Fig9} is illustrated in Fig.~\ref{Fig11}, where all the parameters are the same as in Fig.~\ref{Fig9}. Here only the $e_x$ component is chosen because it is the major component of the field and holds the initial polarization of the incident field.
The singularities denoted by `$S_A$' and `$S_B$',  evolved from the initial singularity (i.e. initial vortex with charge $2$ of the incident field),  play the kernel role in the intensity distribution, thus here we mainly focus on these two singularities.
 From Fig.~\ref{Fig11} we can find:
\begin{itemize}
\item
When $\sigma_c<1$, the $S_A$ and $S_B$ are located symmetrically about the origin along the $x_s$-axis, 
whereas for $\sigma_c\geq 1$, their positions appear on the $y_s$-axis with the same symmetry relation.
\item
As $|\sigma_c|$ approaches $1$, the distance between $S_A$ and $S_B$ becomes shorter.
\end{itemize}
Then, the intensity distributions in Fig.~\ref{Fig9} can be interpreted. 
When $\sigma_c<1$, although the intensity maxima distributes along the $x_s$-axis,  since the locations of the singularities also on $x_s$-axis, the intensity maxima/maximum can stay at the focus [the distance between $S_A$ and $S_B$ is big, see Fig.~\ref{Fig9} (a) and (b)], on $x_s$-axis [the distance between $S_A$ and $S_B$ is short, see Fig.~\ref{Fig9} (c)].
This is also the case for $\sigma_c>1$, and  when $\sigma_c=1.3$ [Fig.~\ref{Fig9} (e)], the distance between $S_A$ and $S_B$ is short [Fig.~\ref{Fig11} (e)] and there are two intensity maxima on $y_s$-axis, while as $\sigma_c$ increases, the distance between $S_A$ and $S_B$ also increases and the intensity pattern show three lobes along $y_s$-axis [Fig.~\ref{Fig9} (g)] and one bright spot at the origin [Fig.~\ref{Fig9} (h)]. Therefore the focal shaping for $n=2$ has been analyzed and interpreted.

Now one may raise another question: why does the vortex split into two phase singularities along the $x_s$-axis at the focal plane for $\sigma_c<1$ or along the $y_s$-axis for $\sigma_c\geq 1$? The answer to this question will be discussed in the following part.

\begin{figure}[ht]
	\centering
	\includegraphics[width=8.0cm]{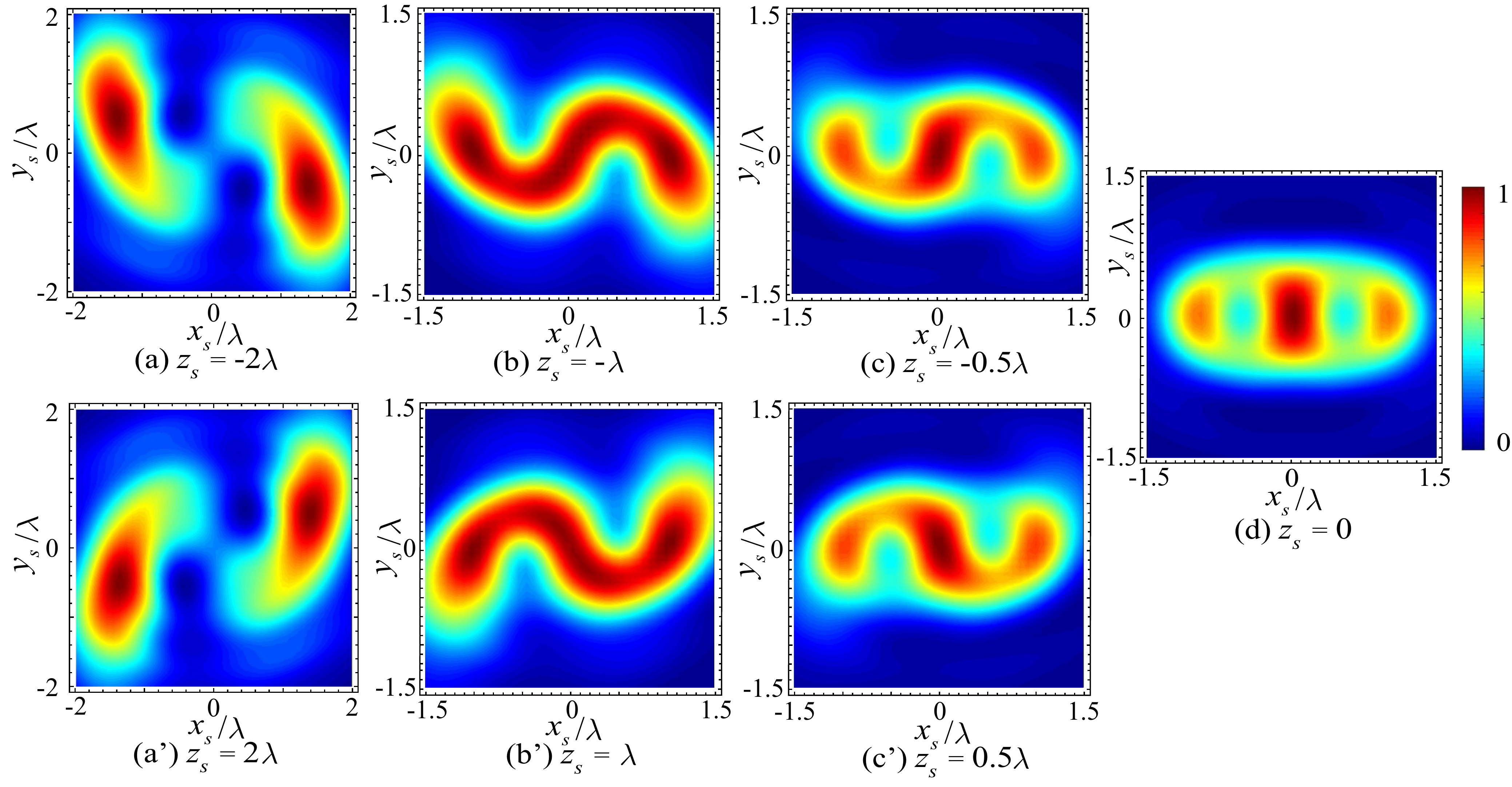}
	\caption{Intensity distribution along the beam propagation.  Here $\alpha=60^\circ$, $\sigma_c=0.3$ and $n=2$. }
	\label{Fig12}
\end{figure}

\begin{figure}[ht]
	\centering
	\includegraphics[width=8.0cm]{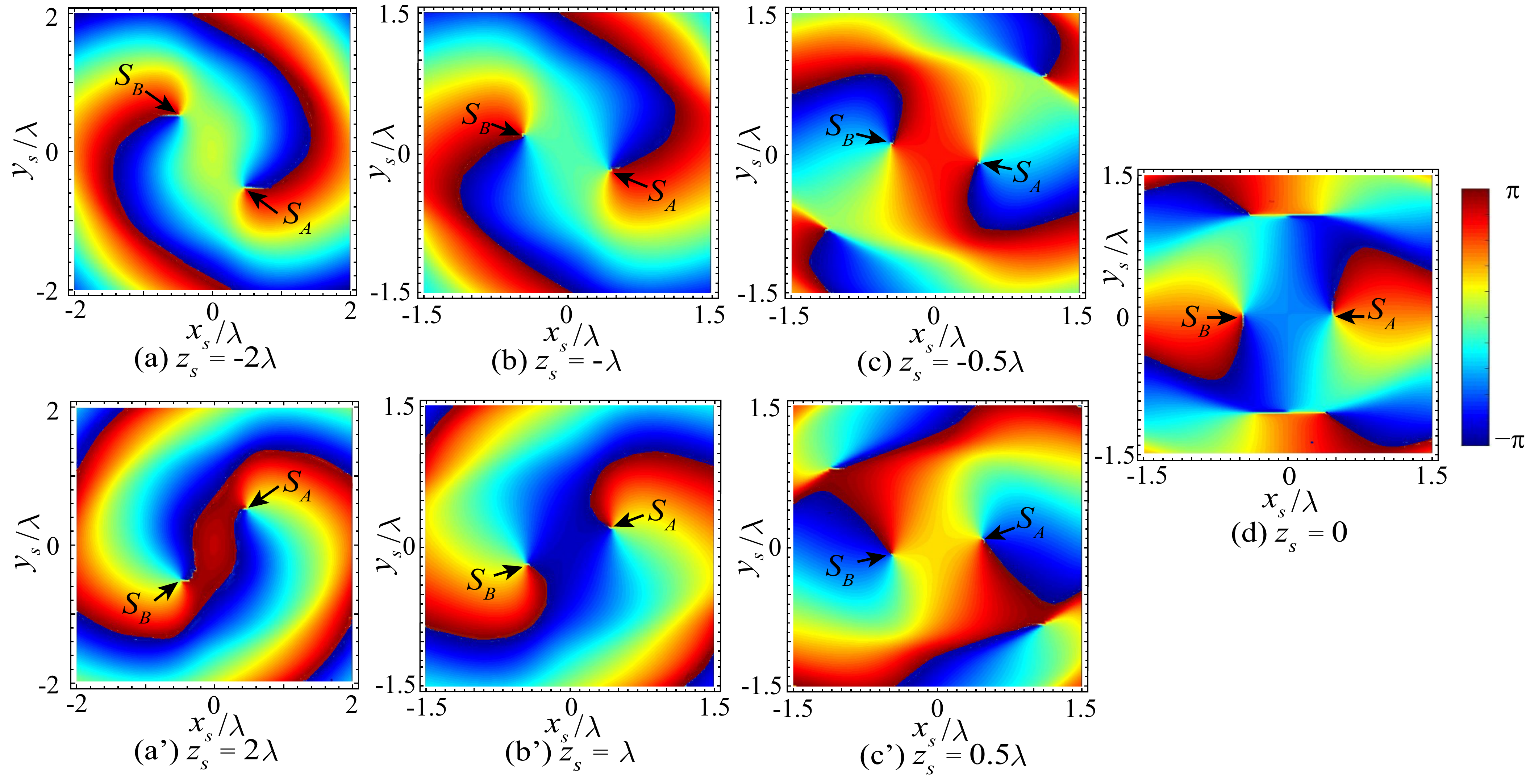}
	\caption{Phase distribution of $e_x$ along the beam propagation. Here $\alpha=60^\circ$, $\sigma_c=0.3$ and $n=2$. }
	\label{Fig13}
\end{figure}

Secondly, we will examine the field distribution along the beam propagation.
The intensity evolution for $n=2$ along the propagation direction is shown in  Fig.~\ref{Fig12}, where $\alpha=60^\circ$, $\sigma_c=0.3$.
One can see that the beam pattern changes from `two-lobe', `S-shape', `one-lobe' to   `S-shape', `two-lobe' again from $z_s=-2\lambda$ to $+2\lambda$, and at the same time the whole pattern  bears a same counter-clockwise rotation as it does in the case of $n=1$ due to the OAM of the beam.
This distinguishing  evolution of intensity pattern is formed mainly by the kernel singularities $S_A$ and $S_B$.
The phase distribution of the $e_x$ component corresponding to the field in Fig.~\ref{Fig12} is illustrated in Fig.~\ref{Fig13} where $S_A$ and $S_B$ are marked out.
It shows that these two singularities also rotate in a counter-clockwise manner, and their appearance disturbs the regular rotation behavior of the intensity distribution, thus enriches the field pattern.
Actually the phase structure of the field on one side of the focus can be predicted by the phase on the other side from the symmetry relation of the focused field, Eqs. (\ref{es1})-(\ref{es3}).
According to these equations, the phase functions of $e_x$, $e_y$, $e_z$ satisfy:
 \begin{align}\label{eph1}
&-\Phi_x(\rho_{s},\phi_{s},z_{s})=\pi+\Phi_x(\rho_{s},\pi-\phi_{s},-z_{s}),  \\
&-\Phi_y(\rho_{s},\phi_{s},z_{s})=\Phi_y(\rho_{s},\pi-\phi_{s},-z_{s}), \\
&-\Phi_z(\rho_{s},\phi_{s},z_{s})=\pi+\Phi_z(\rho_{s},\pi-\phi_{s},-z_{s}), 
\end{align}
and from Eq.~(\ref{eph1}) the phase of $e_x$ on the $-z_s$ side can be drawn from that on the $+z_s$.
The symmetry relation of the phase actually also implies the rotation of the vortex.  

\begin{figure}[ht]
	\centering
	\includegraphics[width=7.0cm]{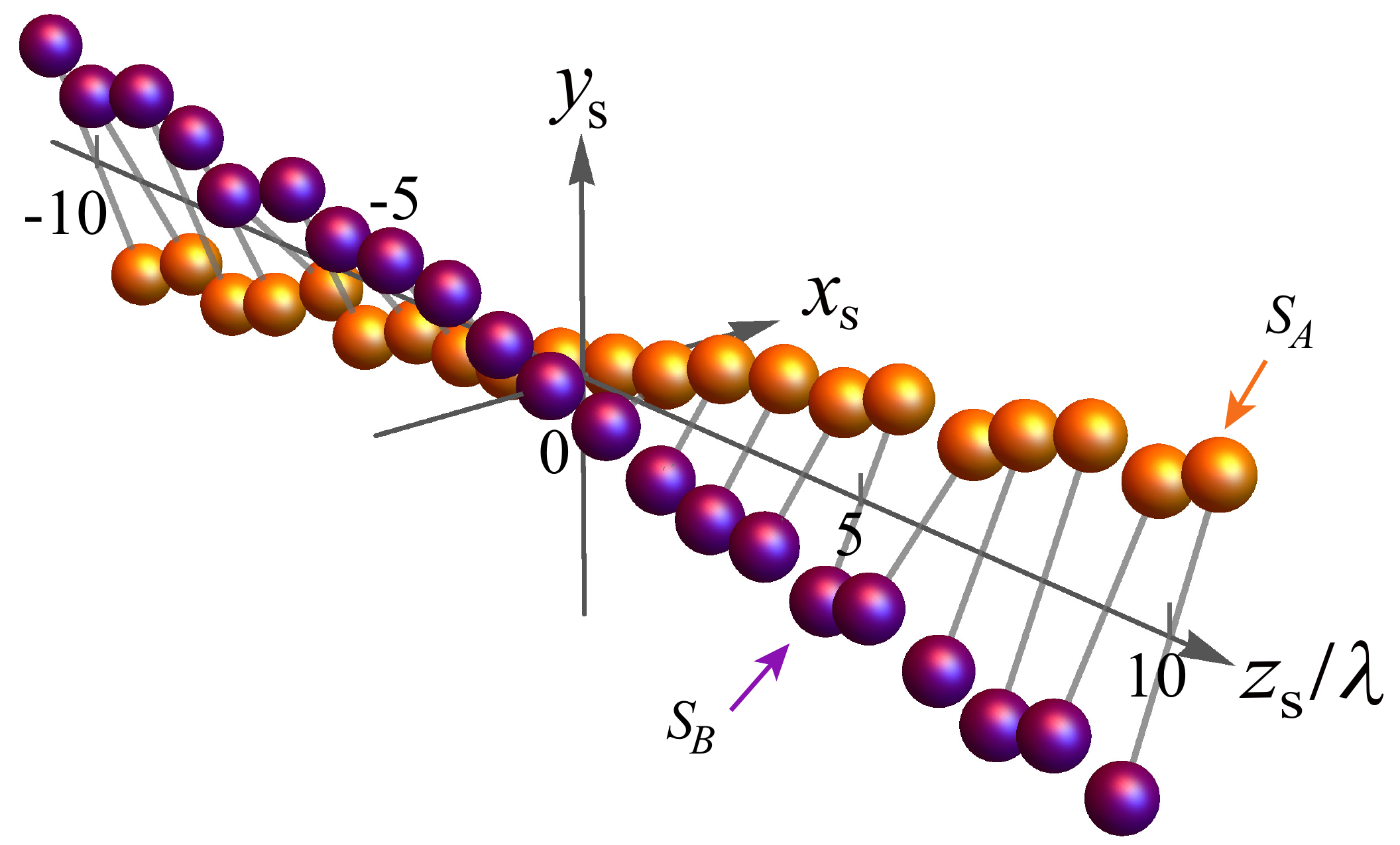}
	\caption{Positions of two vortices with $z_s$ from $-10\lambda$ to $10\lambda$ (see \textbf{Video 2}). Here $\alpha=60^\circ$, $\sigma_c=0.3$ and $n=2$. }
	\label{Fig14}
\end{figure}

\begin{figure}[ht]
	\centering
	\includegraphics[width=8.0cm]{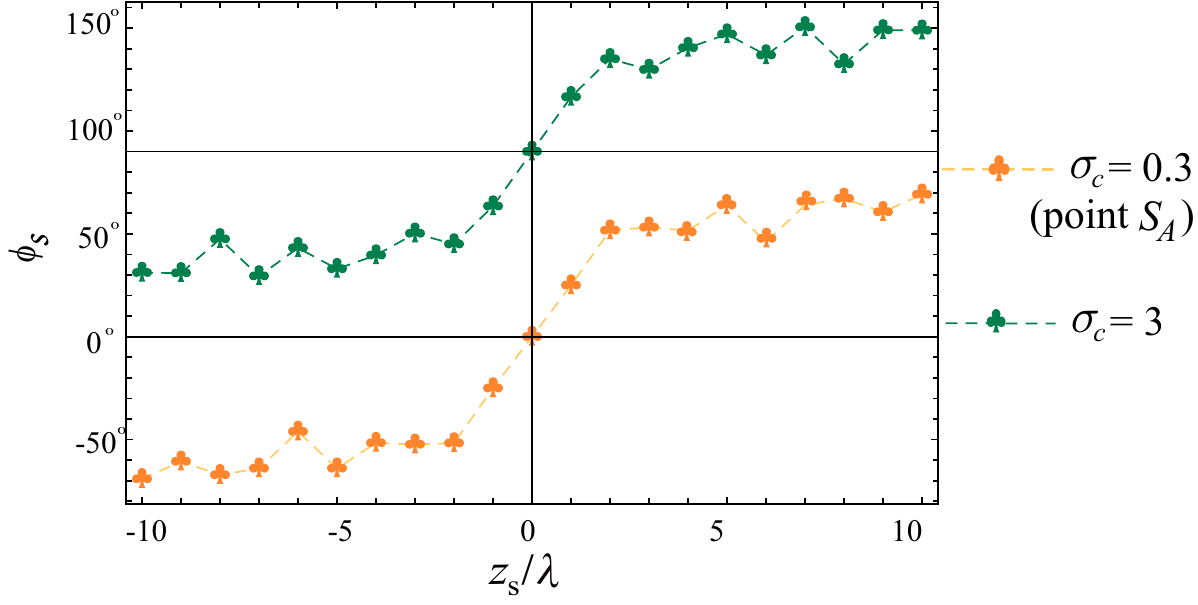}
	\caption{Azimuthal angle of the vortex along the propagation direction for the incident beam of charge $+2$. Here $\alpha=60^\circ$. }
	\label{Fig15}
\end{figure}

\begin{figure}[ht]
	\centering
	\includegraphics[width=8.0cm]{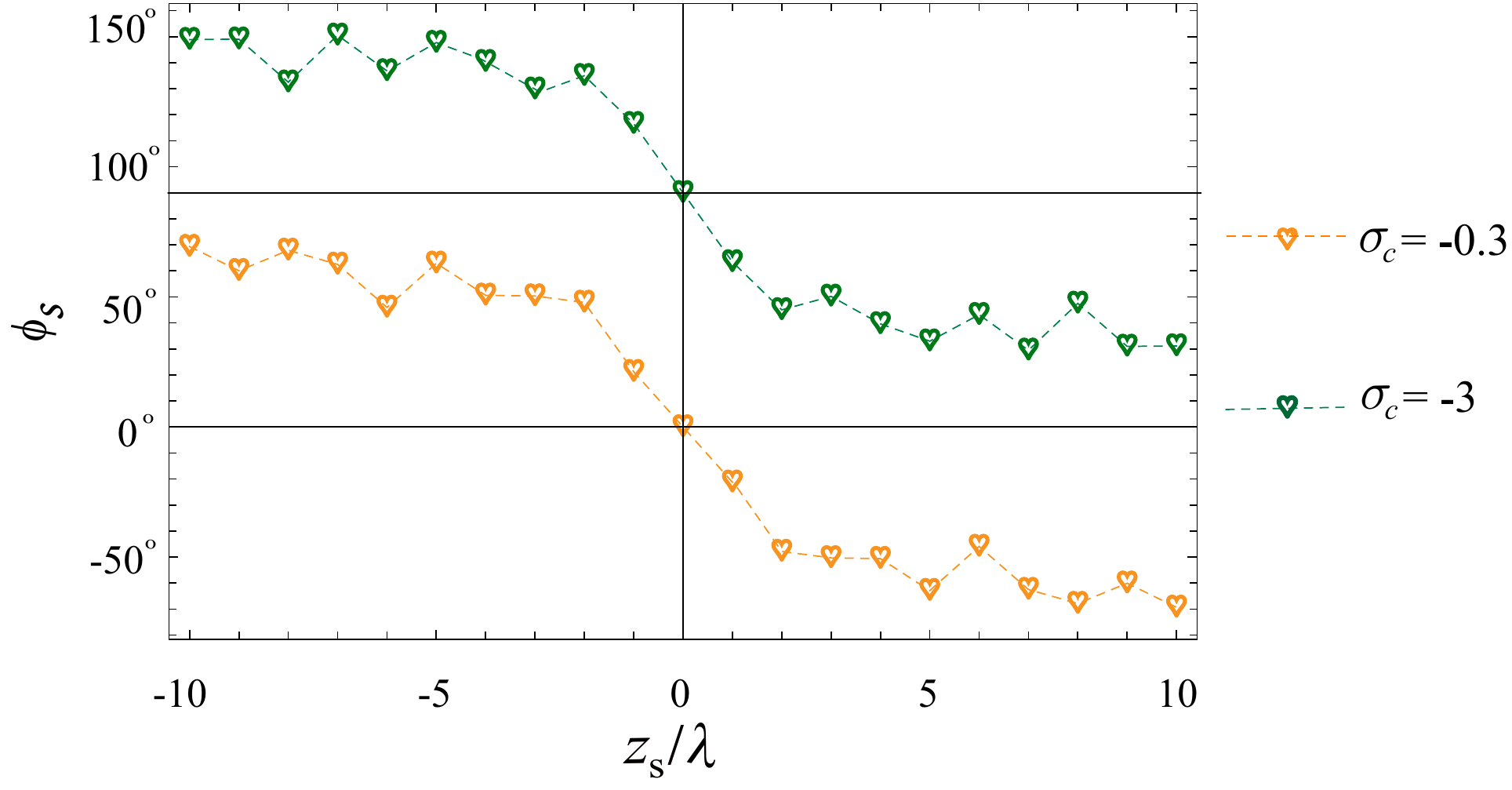}
	\caption{Azimuthal angle of the vortex along the propagation direction for the incident beam of charge $-2$. Here $\alpha=60^\circ$. }
	\label{Fig16}
\end{figure}

So far we have explained the distinguishing beam shaping for $n=2$ from the rotation behavior of the phase singularities.
Now it will be of interest to get a further understanding of the propagation property of these singularities. 
The positions of the kernel singularities, $S_A$ and $S_B$ in the transverse planes with $z_s$ from $-10\lambda$ to $10\lambda$ are displayed in Fig.~\ref{Fig14}, where the orange ball and the purple ball represent the positions of $S_A$ and $S_B$ respectively (the parameters here are the same as in Fig. \ref{Fig13}).
From this figure, the counter-clockwise rotation is seen more directly.
 It also can be observed that the  $S_A$ and $S_B$  behave more  regularly than the intensity maxima for $n=2$, 
which is because of the topological stable of the phase singularities \cite{s2001,gbur2017}.
The rotation angle of point $S_A$ is displayed in Fig.~\ref{Fig15}, and we can see that the azimuthal angle  is increased from $-70^\circ$ to $70^\circ$ with $z_s$ from $-10\lambda$ to $10\lambda$ for $\sigma_c=0.3$. 
The azimuthal angle of the vortex in the case of $\sigma_c=3$ is also displayed and 
it shows that the vortex also rotates counter-clockwise from $30^\circ$ to $140^\circ$.
As we discussed in section 2, the rotation orientation depends on the direction of the OAM of the incident beam.
In order to illustrate this point, the azimuthal rotation of the vortex for $\sigma_c<0$ is also shown in Fig.~\ref{Fig16}, 
where one can see that the vortex with negative charge will rotate in a clockwise manner along the propagation direction.

Furthermore,
Figs.~\ref{Fig15} and \ref{Fig16}, as well as Eq. (\ref{eq5}) also indicate that if $z_s$ goes to $-\infty$, the two vortices will be on the $y_s$-axis ($\phi_s=\pm 90^\circ$) for $|\sigma_c|<1$, and on the $x_s$-axis ($\phi_s=0^\circ$, $180^\circ$)  for $|\sigma_c|>1$.
It means that as the beam propagates from $-\infty$ to the focus the vortices rotate $90^\circ$ transversely from the $y_s$-axis to the $x_s$-axis for $|\sigma_c|<1$, whereas they rotate from the $x_s$-axis to the $y_s$-axis for $|\sigma_c|>1$.
This behavior is generated by two factors:
the phase gradient of the X-type vortex and the propagation property of the off-axis vortex. 
From Eq.~(\ref{eq4}), one can get that 
\begin{align}
|\nabla\Phi| = \left \{
\begin{array}{cl}
|n\sigma_c|, &  \mbox{on $x$-axis}  \\
|n/\sigma_c|, & \mbox{on $y$-axis}
\end{array}
\right.
\end{align}
Thus, when $|\sigma_c|<1$, for instance $|\sigma_c|=0.3$, $|\nabla\Phi|\approx 6.67$ on the $y$-axis and $|\nabla\Phi|=0.6$ on the $x$-axis.
Since $6.67>0.6$ and the `nongeneric' characteristic of the higher order vortex, the vortex with topological charge $\pm 2$ is easy to degenerate into $2$ vortices along the $y_s$-axis as the beam passes through the lens.
Similarly, if $|\sigma_c|>1$, the vortex will split into $2$ vortices located on the $x_s$-axis.
So these two split vortices become off-axis at the `beginning' of the focusing.
 It has been demonstrated that the off-axis vortices will experience a $90^\circ$-rotation to the focus and for the vortices with the same sign their rotation directions are the same (i.e. clockwise or counter-clockwise) \cite{XY17}.
Therefore, for $|\sigma_c|=0.3$, the singularities $S_A$ and $S_B$ will rotate from near the $y_s$-axis to the $x_s$-axis as $z_s$ from $-10\lambda$ to $0$;
while for $|\sigma_c|=3$, they rotate from near the $x_s$-axis to the $y_s$-axis.
Thus the behaviors of the vortices in Figs.~\ref{Fig15} and \ref{Fig16} have been explained.
Then we will come to the question raised before:
what is the reason for the kernel phase singularities located along the $x_s$-axis for $\sigma_c<1$ and along the $y_s$-axis for $\sigma_c\geq 1$?
Now the answer is clear:
because of the phase gradient of the X-type vortex and the $90^\circ$ rotation of the off-axis vortex, at the focal plane the split singularities ($S_A$ and $S_B$) located on the $x_s$-axis for $|\sigma_c|<1$ and on the $y_s$-axis for $|\sigma_c|>1$.
For $|\sigma_c|=1$, the location of $S_A$ and $S_B$ on  $y_s$-axis  is caused by another factor: the $x$-polarization of the incident beam. It is easy to demonstrate that if the beam is $y$-polarized at the entrance pupil, the two kernel singularities will be seen on the $x_s$-axis at the focal plane.

Now we can have a summary: 
when $n=2$, the phase structure (i.e. the phase gradient and the topological charge) of the X-type vortex has a crucial role in the beam shaping.
The rotating property of the kernel singularities, such as the rotation direction, location of the singularities and their relative distance, determines the spatial intensity patterns on both the focal plane and the 3D propagating space.
In addition, at the focal plane the kernel singularities are distributed symmetrically along the $x_s$-axis for $|\sigma_c|<1$, while they are  along the $y_s$-axis symmetrically for $|\sigma_c|\geq 1$.

\begin{figure}[ht]
	\centering
	\includegraphics[width=8.0cm]{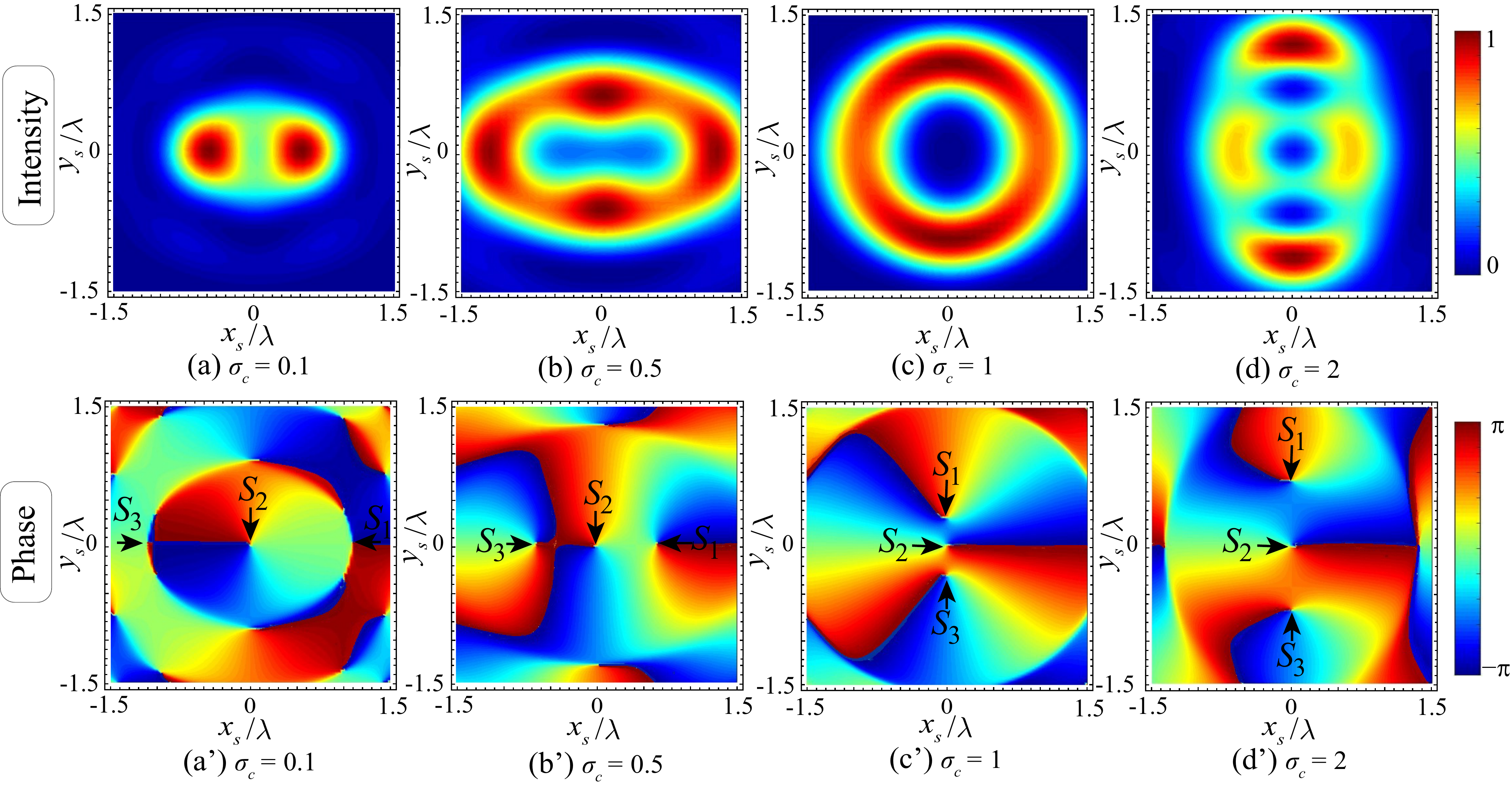}
	\caption{The field distribution at the focal plane for the X-type vortex with $n=3$. The intensity pattern is shown in the first line, and the corresponding phase of $e_x$ component is shown in the second line.  Here $\alpha=60^\circ$. }
	\label{Fig17}
\end{figure}

\begin{figure}[ht]
	\centering
	\includegraphics[width=8.0cm]{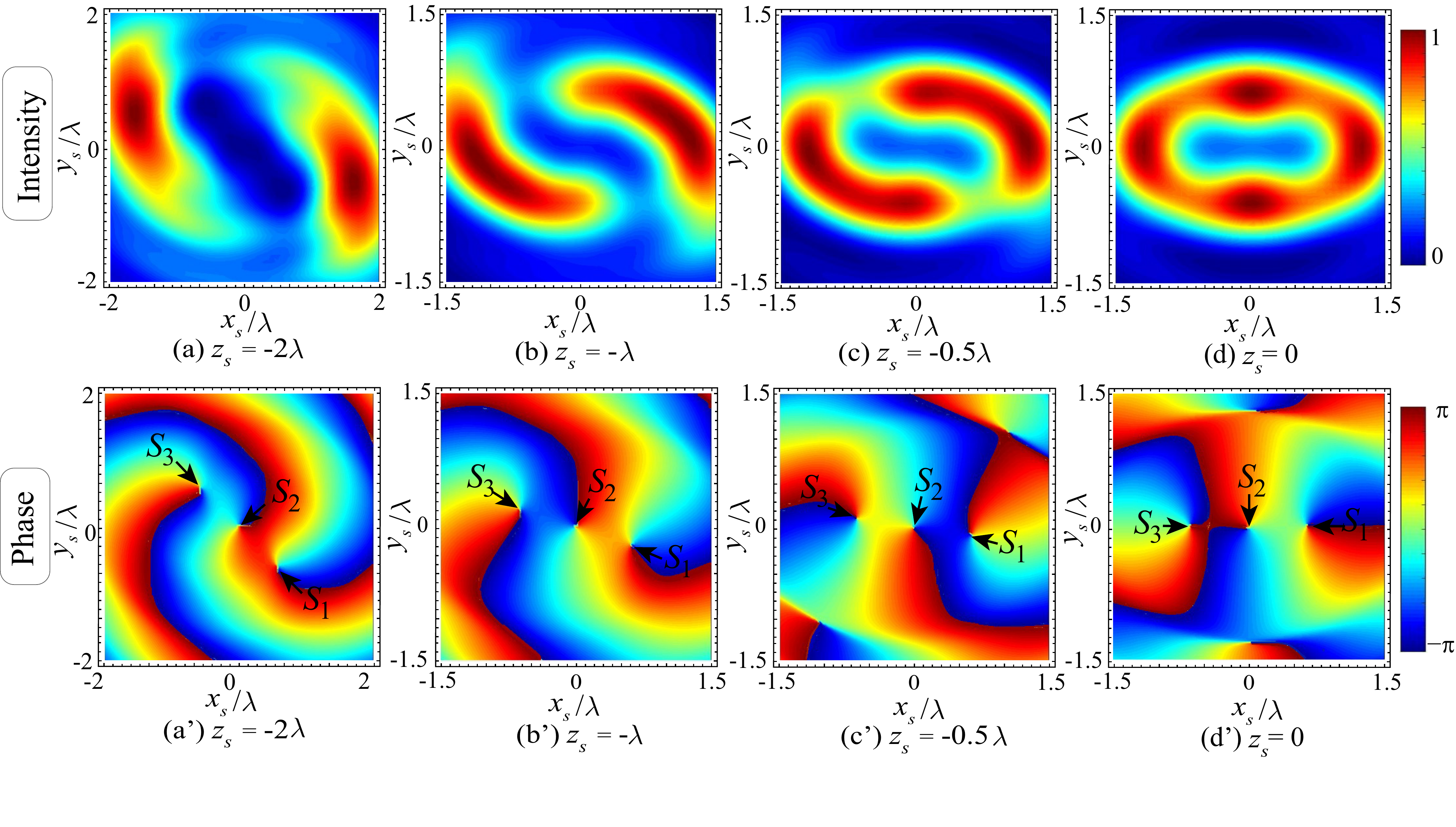}
	\caption{The field distribution with the beam propagation for the X-type vortex with $n=3$. The intensity pattern is shown in the first line, and the corresponding phase of $e_x$ component is shown in the second line.  Here $\alpha=60^\circ$, $\sigma_c=0.5$. }
	\label{Fig18}
\end{figure}

The above conclusion can also be generalized to any case for $n>1$.
As an example, in Fig.~\ref{Fig17} the total intensity distribution and the corresponding phase of $e_x$ component at the focal plane for $n=3$ are illustrated, where the semi-aperture angle $\alpha=60^\circ$, $\sigma_c=0.1, 0.5, 1$,$2$, and $n=3$.
As one can see, the intensity patterns can be `two-lobe' along different axes, `four-lobe' and `doughnut shaped', which mainly depends on the locations of three kernel singularities, $S_1$, $S_2$ and $S_3$.
It is also clear that $S_1$, $S_2$ and $S_3$ are located along the $x_s$-axis for $\sigma_c=0.1$ and $0.5$, while they have their positions on the $y_s$-axis for $\sigma_c=1$ and $2$.
The field distribution along the propagation direction is shown in Fig. \ref{Fig18},
from which one can see that the three kernel singularities obey the rule summarized above:
they rotate in a counter-clockwise manner along the propagation direction for $\sigma_c=0.5$.
Figs.~\ref{Fig17} and \ref{Fig18} also show that the positions of the kernel phase singularities, their distance and rotation behavior  shape the beam patterns at the focal plane and in the 3D propagation space. 

\section{Conclusions}
A new type of noncanonical vortex, the X-type vortex is defined and its effects on shaping the  intensity spatial distribution of a strongly focused field are discussed. 
We find that,
firstly there exists a `switch effect' of the X-type vortex,  which shows that when the anisotropic parameter $\sigma_c$ of the vortex passes through the switch value $\sigma^{sw}_c$ ($\sigma^{sw}_c\approx \pm 1$), the intensity maxima at the focal plane will `switch' from one transverse axis (for instance the $x_s$-axis) to another (the $y_s$-axis).
Even a small change in $\sigma_c$ near $\sigma^{sw}_c$, which may occur easily in generating a canonical vortex (i.e. $|\sigma_c=1|$), can lead to a $90^\circ$ variation in the beam pattern.
This may have implications for generation of optical vortices in experiments.
Secondly, the inconstant phase gradient of the X-type vortex  brings a `rotation' structure of the beam in the 3D focal region, which physically is caused by the joint effect of the OAM and the anisotropic phase gradient of the vortex.
This finding implies by changing the sign of $\sigma_c$ (i.e. the direction of the OAM) and magnitude of $\sigma_c$ can control the rotation direction and the position of the twisting strips respectively.
Thirdly, when $|n|>1$, the `nongeneric' characteristic of the X-type vortex exerts  a strong influence on tailoring rich field patterns in focal plane and diverse rotation strips of the beam intensity in 3D space.
The physical mechanism underlying this phenomenon can be explained by the topological propagation theory of the vortices, that also suggests a new path to shape beams in 2D and 3D optical fields.
Our findings offer a new vision for
the fundamental properties of optical vortices, and are
expected to attract further interest to applications of optical singularities.

\section*{Acknowledgement}
    The work was supported by National Natural Science Foundation of
	China (NSFC) (No. 11974281, 12104283), Natural Science Basic Research Plan in Shaanxi Province of China (No. 2020JM-116), Fundamental Research Funds for the Central Universities (No. GK202103021).

\section*{Compliance with Ethical Standards}
{\bf Conflict of Interest:} The authors declare that there are no conflicts of interest related to this article. 


\bibliographystyle{iopart-num}

\bibliography{XType}

\end{document}